\documentclass[twocolumn]{aastex61}
\usepackage{amsmath}
\usepackage{color}\hypersetup{linkcolor=red,citecolor=green,filecolor=cyan,urlcolor=magenta}
\received{}
\revised{}
\accepted{}
\submitjournal{ApJ}
\begin{document}

\title{Relativistic spacecraft propelled by directed energy}

\correspondingauthor{Neeraj Kulkarni}
\email{neeraj.k@me.com}

\author{Neeraj Kulkarni}
\affil{Department of Physics \\ University of California\\ Santa Barbara, CA 93106, USA}

\author{Philip Lubin}
\affil{Department of Physics \\ University of California\\ Santa Barbara, CA 93106, USA}

\author{Qicheng Zhang}
\affil{Department of Physics \\ University of California\\ Santa Barbara, CA 93106, USA}

\begin{abstract}

Achieving relativistic flight to enable extrasolar exploration is one of the dreams of humanity and the long term goal of our NASA Starlight program. We derive a fully relativistic solution for the motion of a spacecraft propelled by radiation pressure from a directed energy system. Depending on the system parameters, low mass spacecraft can achieve relativistic speeds; thereby enabling interstellar exploration. The diffraction of the directed energy system plays an important role and limits the maximum speed of the spacecraft. We consider 'photon recycling' as a possible method to achieving higher speeds. We also discuss recent claims that our previous work on this topic is incorrect and show that these claims arise from an improper treatment of causality. 
\end{abstract}


\section{Introduction}
In the past half-century, we have seen incredible advances in space exploration through chemically powered rockets. However, rockets powered by combustion are extremely inefficient and need to carry massive amounts of fuel. For example, suppose we want to propel a single proton with the entire mass of the universe ($\sim 10^{56}$ g) converted to chemical propellant. Treating this system as an infinitely staged rocket with a typical ejection speed of $\sim 4$ km s$^{-1}$, the final speed of the proton is:

\begin{equation}
{v_f} \sim 4\text{ km\,s}^{-1}\ln \left( {\frac{{{m_{universe}}}}{{{m_{proton}}}}} \right) \sim 0.002c
\end{equation}

Chemical fuel cannot accelerate even a single proton to relativistic speeds and thus is hopeless for interstellar flight in a reasonable flight time. If we want to explore distances far beyond our solar system, we need to adopt a radically different form of propulsion. All other known methods of propulsion including ion engines, nuclear thermal, fission fragment and fusion fail due their low inherent energy efficiency ($\varepsilon = $ energy released per rest mass energy) with even fusion having $\varepsilon < 0.01$. For reference, chemical rockets have $\varepsilon<10^{-10}$. Even antimatter annihilation engines would have significant difficulty due to the large secondary masses required for confinement and reaction \citep{lubin2016roadmap}. 

In our NASA Starlight program and it's derivative, the Breakthrough Starshot program, we propose using directed energy (DE) photon propulsion. Unlike all other propulsion technologies, the DE system does not 'fly'; it is used in a beamed energy mode to remotely propel the spacecraft (described in detail in \citep{lubin2016roadmap}). The energy emitted by the DE system reflects off of a 'sail'. This imparts momentum to the attached spacecraft propelling it forward. DE propulsion is \textit{highly versatile} and can be used to propel spacecraft with a wide range of masses. 

Due to the advances and exponential growth of both electronics and photonics, highly functional spacecraft can be greatly miniaturized - in some cases to a gram-scale. These gram-scale spacecrafts can achieve relativistic speeds $\sim 0.2c$ allowing exoplanet missions in flight times of  $\sim$ 20 years. Such a program would be a radical transformation from existing propulsion technologies and would enable missions that are currently not possible. While this is a long term program that is multi-decade in it's scope, it would be transformational in nature.
One can envision an armada of small spacecrafts traveling at relativistic speeds to explore the nearby stars and exoplanets; thereby extending humanity's reach far beyond the solar system.

There has been a variety of work on the general problem of a mirror/sail propelled by radiation pressure. 
This problem was first imagined by \citet{einstein1905elektrodynamik} and subsequently by \citet{marx1966interstellar}, \citet{redding1967interstellar} and \citet{forward1984roundtrip}. Our 
work on this topic consists of:
\citet{bible2013relativistic}, \citet{lubin2015directed} and \citet{lubin2016roadmap} which contains a comprehensive review of our work till date. Most recently, \citet{kipping} has also published a paper on this topic. In this paper, we begin by deriving a relativistic solution for the velocity of the spacecraft using the conservation of 4-momentum in Section \ref{sect:Methodology}. In doing so, we dispel claims made in \citet{kipping} stating that our previous work on this subject is incorrect. Section \ref{sect:eff} contains a general discussion of the efficiency of DE propulsion. In Section \ref{sect:diffraction}, we discuss diffraction effects of the DE system and how this results in a terminal speed for the spacecraft. Using this methodology, we find the spacecraft velocity for various parameters (spacecraft mass, thickness of the sail, power of the DE system etc.) in table \ref{missionscenarios}. In particular we find that a 1 g spacecraft propelled by a 100 GW (10 km) DE system can reach speeds $\sim 0.2c$. A preliminary model of 'photon recycling' is discussed as a possible method for increasing the speed of the spacecraft in Section \ref{sect:pr}. Finally, we end with some concluding remarks in Section \ref{sect:conclusion}.

\section{Derivation of relativistic solution}
\label{sect:Methodology} 
Consider an DE system emitting a stream of photons which strike a moving sail. Momentum is transferred to the spacecraft through collisions between photons and the sail. In any single collision, 4-momentum is conserved and can analyzed in a specific, but arbitrary reference frame. In such a frame frame, the incoming photon has an energy $E_0$ and strikes a sail with mass $m$ and momentum $p_0$.

4-momentum conservation tells us:

\begin{equation}
\tilde p_0^\mu  + p_0^\mu  = \tilde p_f^\mu  + p_f^\mu 
\end{equation}

We use notation where the tilde corresponds to the photon. Written explicitly in terms of components:

\begin{equation}\label{simeqn}
     \begin{bmatrix}
    {\left| {{E_0}} \right| + \sqrt {{m^2} + p_0^2} } \\
    {{E_0} + {p_0}} \\
    \end{bmatrix} =      
    \begin{bmatrix}
    {\left| {{E_f}} \right| + \sqrt {{m^2} + p_f^2} } \\
    {{E_f} + {p_f}} \\
    \end{bmatrix}	
\end{equation}

We set $c=1$ and restore factors of $c$ when useful. The above system of equations can be solved for $p_f$ and $E_f$.

\begin{equation}\label{messysoln}
\begin{split}
p_f\left(p_0 , E_0 \right) &= \left({4{E_0}^2 + 4{E_0}{p_0} - {m^2}}\right)^{-1} 
\big(4{E_0}^3 \\+& 2{E_0}\left( {\sqrt {{E_0}^2\left( {{m^2} + {p_0}^2} \right)}  - {m^2} + {p_0}^2} \right) \\+& {p_0}\left( {2\sqrt {{E_0}^2\left( {{m^2} + {p_0}^2} \right)}  - {m^2}} \right) + 6{E_0}^2{p_0}\big)
\end{split}
\end{equation}

Similarly, solving for $E_f$ produces an equally unwieldy expression. So far, this analysis is indistinguishable from a Compton scattering calculation in 1 spatial dimension. In this single collision, the change in the spacecraft's 4-momentum will be: $\Delta {p^\mu } = p_f^\mu  - p_0^\mu $. In general,  $\Delta {p^\mu }$ will be some function of the spacecraft's current momentum and the incident energy of the photon. 

Now consider an entire stream of photons with energy $E_0$ striking a perfectly reflecting sail. Each of these photons will impart energy and momentum to the sail. In the limit that the energy of any individual photon is small compared to the rest energy of the sail, the energy transferred by each individual photon is negligible and the collisions can be smoothed over. If we assume that the DE system is massive enough to have a negligible recoil when firing photons, we can approximate the DE system as remaining in an inertial frame; thereupon, we analyze the situation in the DE system's rest frame. 

In it's frame, the DE system of power $P$ emits photons of energy $E_0$ at a rate $\Gamma$ ($\Gamma$ has units of inverse time). Due to the relative motion between the sail and DE system, the rate at which photons are emitted from the DE system is \textit{not} rate at which photons strike the sail. Let $dN$ be the number of photons that strike the sail in $dt$, where $dt$ refers to a small proper time elapsed in the DE system's frame. 

\begin{equation}\label{smoothing}
dN = \left( {1 - v} \right)\Gamma dt = \left( {1 - \frac{p}{{\sqrt {{m^2} + {p^2}} }}} \right)\Gamma dt
\end{equation}

Smoothing over individual collisions in such a manner gives the change in 4-momentum of the spacecraft in time $dt$:

\begin{equation}\label{fundamentaleqn}
\frac{{d{p^\mu }}}{{dt}} = \Gamma \Delta {p^\mu }\left( {1 - \frac{p}{{\sqrt {{m^2} + {p^2}} }}} \right)
\end{equation}
For clarity, we define a dimensionless photon energy $\epsilon = {E_0}/{mc^2}$. We apply this analysis to systems like our NASA Starlight program, the Breakthrough Starshot program \citep{merali2016shooting} and DE-STAR \citep{hughes2013star}. These involve propelling gram-scale spacecraft with $\sim 100$ GW DE systems which shoot out $\sim 1$ eV photons. For these parameters: $
\epsilon \sim {1 \text{ eV}}/{\left(1\text{ gram}\right) c^2} \sim 10^{-33}$. 

Given an initial position and velocity, equation \ref{fundamentaleqn} can be integrated numerically to solve for the spacecraft's motion as seen in the DE system's frame. This derivation involved:

\begin{enumerate}
	\item \textit{conservation of 4-momentum applied to a single collision event (in an arbitrary frame)}
	
	\item \textit{smoothing over multiple collisions to get a continuous force (as seen in the DE system's frame)}
\end{enumerate}

Various phenomenon that emerge in special relativity (i.e. length contraction, time dilation, Doppler shifts) are implicitly included in our assumptions. For instance, calculating $\left|E_{f}\right|$ from equation \ref{simeqn} to first order in $\epsilon$ gives: 

\begin{equation}\label{doppler}
\left| {{E_f}} \right| = \left( {\frac{{1 - \beta}}{{1 + \beta}}} \right)\left| {{E_i}} \right|
\end{equation}

As expected, the reflected photon encounters 2 Doppler shift factors; first when it is being absorbed by the receding sail and second when it is being emitted by the receding sail. In this sense, our derivation is extremely general.

Our second assumption is valid as long as the timescale between individual collisions ($t_{coll}^{-1} \equiv (1-v/c)\Gamma$) is much shorter than the timescale for the spacecraft to appreciably change it's velocity ($t_{rel} \equiv mc^2/P$). Even if we assume extremely relativistic speeds of $0.99c$, $t_{coll}/t_{rel} \sim 10^{-31}$.

Since we're interested in the soft photon limit where $\epsilon \ll 1$, from here on, we'll primarily work to first order in $\epsilon$.

\begin{equation}
\begin{split}
\Delta p^1 =& {p_f}\left( {p,\epsilon } \right) - p \\=& 2\left( {\frac{{{m^2} + {p^2} - p\sqrt {{m^2} + {p^2}} }}{m}} \right)\epsilon + \mathcal{O}(\epsilon^2)
\end{split}
\end{equation}

Using $P = E_0\Gamma$ and $v$ as the instantaneous velocity of the spacecraft, the spatial component of equation \ref{fundamentaleqn} becomes:

\begin{equation}\label{1stepaway}
\frac{{d{p^1}}}{{dt}} = \frac{d}{{dt}}\left( {\gamma mv} \right) = 2P\left( {\frac{{1 - v}}{{1 + v}}} \right)
\end{equation}

Restoring factors of $c$ and defining $\beta = v/c$, equation \ref{1stepaway} can be written as:

\begin{equation}\label{simplediffeqnbeta}
\dot \beta  = \frac{{2P}}{{m{c^2}{\gamma ^3}}}\left( {\frac{{1 - \beta }}{{1 + \beta }}} \right)
\end{equation}

In the limit of $\epsilon \ll 1$, equation \ref{simplediffeqnbeta} can be integrated to give an analytic expression relating the time $t$ taken to get to a given $\beta$.

\begin{equation}\label{vclosedformt}
t = \frac{t_{rel}}{6}\left[ {\frac{{\left( {1 + \beta } \right)\left( {2 - \beta } \right)\gamma }}{{\left( {1 - \beta } \right)}} - 2} \right]
\end{equation}

In the non-relativistic limit, equation \ref{vclosedformt} is consistent with the Newtonian impulse due to radiation pressure: $F_{rad}t = mv$. Where $F_{rad} \equiv 2P/c$. 

\begin{figure}
	\includegraphics[width=8cm]{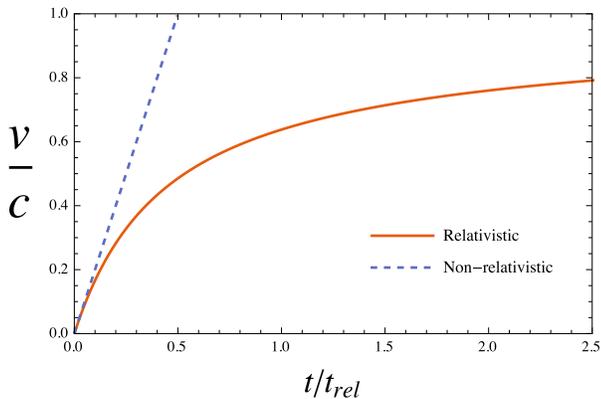}
	\caption{A plot of equation \ref{vclosedformt} showing the velocity of the spacecraft as a function of time (as seen in the DE system's frame). By a time $t=t_{rel}$, relativistic effects become very significant. We assume continued illumination by the DE system and that the sail is perfectly reflecting at all wavelengths. Diffraction effects are ignored here but discussed in section \ref{sect:diffraction}}
	\label{simpletraj}
\end{figure}

\subsection{Resolving recent claims of inconsistency}

There is a variety of literature on the topic of relativistic travel using DE propulsion. \citet{lubin2016roadmap} states the relativistic solution which is derived in detail here.  Since then, there has been a recent paper \citep{kipping} claiming that our solution is wrong. In particular, Figure 4 of \citep{kipping} and the subsequent text claims that their result "significantly diverges from (ours), predicting  10.8$\%$ more energy needed to reach $0.2c$". \citep{kipping} claims that this discrepancy is due to our "infinite mass sail approximation". 

From our analysis, it should be clear that the claim in \citep{kipping} is incorrect. Corrections to equation \ref{simplediffeqnbeta} will be second order in $\epsilon$ and therefore will be of order 1 part in $10^{66}$ for our system. As an extreme example, we can treat our spacecraft as a \textit{single electron}; in which case the problem reduces to Compton scattering \citep{PhysRev.21.483}. The change in the photon's wavelength is given by:

\begin{equation}\label{comptoneq}
\Delta \lambda  = \frac{h}{{{m_e}c}}\left( {1 - \cos \theta } \right)
\end{equation}

For our system with photons $\sim 1$ eV, $\Delta\lambda/\lambda = 2h/m_ec\lambda\sim 10^{-6}$. This is a small effect for a single electron; let alone an entire spacecraft.

Like most paradoxes in relativity, Kipping's apparent discrepancy of $10\%$ arises from an improper treatment of causality.   

\begin{figure}
	\centering
	\includegraphics[width=8cm]{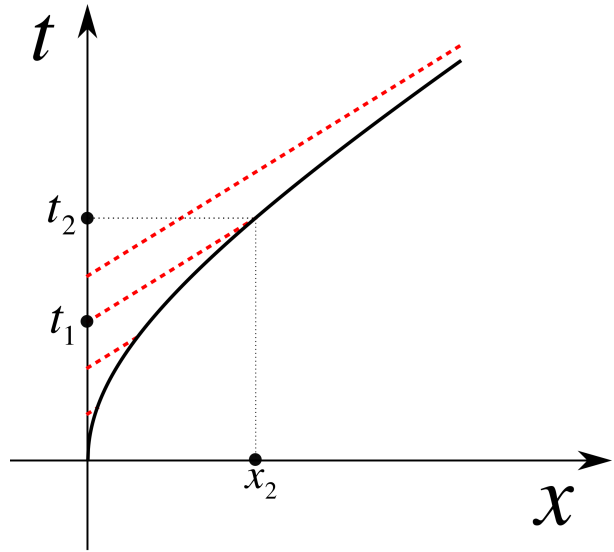}
	\caption{A spacetime diagram of the spacecraft, DE system. The DE system is located at $x=0$ and continuously emits photons shown by the dashed red lines. The trajectory of the spacecraft is represented by the solid black line. Due to the finite speed of light, the spacecraft at time $t_2$ can only be affected by photons that were emitted before a time $t_1$.}
	\label{kippingcausal}
\end{figure}

\citep{kipping} uses the result we quote in equation \ref{vclosedformt} to find the time it takes to get to $0.2c$. This time is multiplied by the power and \citep{kipping} claims this value differs from his by $10\%$. As illustrated by the space-time diagram in figure \ref{kippingcausal}, Kipping's interpretation is misguided. Photons do not strike the sail instantly upon emission. 

Suppose at time $t_2$, the spacecraft is at position $x_2$. It can only be affected by light emitted before a time of $t_1$. From  equation \ref{vclosedformt}, $t_2$ is the time it takes the spacecraft to reach a speed of $0.2c$. $t_2 = 0.126 t_{rel}$. Assuming $x\left(0\right) = 0$, $x_2 = x\left(t_2\right)$, numerically integrating equation \ref{simplediffeqnbeta} gives $x\left(t\right)$. Finally, consider a light ray moving in the positive $x$ direction passing through $(t_2,x_2)$. Tracing this light ray back to when it was emitted by the DE system at $x=0$ gives: $t_1 = 0.112 t_{rel}$. We can now define an 'error':

\begin{equation}
error \equiv \frac{{{t_2} - {t_1}}}{{{t_2}}} \approx 10.8\%
\end{equation} 

Since the total energy emitted by the DE system scales linearly with time, we see the origin of Kipping's false claim: "\citep{lubin2016roadmap} predicts $\sim 10.8\%$ more energy needed to reach 0.2c". The key points of this discussion are:

\begin{itemize}
	\item \citep{kipping} claims that the solution we derive here is incorrect due to our "infinite mass sail approximation". 
	\item We show this claim to be \textit{incorrect}. 
	\item The discrepancy arises because \citep{kipping} assumes that photons strike the sail instantly upon emission. Since light has a finite speed, this cannot be the case.
	\item By including the flight time of the photons, we explicitly derive the apparent discrepancy between \citep{kipping} and our results.  
\end{itemize}

We emphasize that this discrepancy is \textit{not} from finite mass effects of the spacecraft, but from a \textit{lack of causality} in the calculations in \citep{kipping}.

\section{Efficiency of energy transfer}
\label{sect:eff}

As photons reflect off the sail, they get redshifted; thereby transferring energy to the spacecraft. To first order in $\epsilon$, equation \ref{doppler} gives the energy transferred in a collision.

\begin{equation}
\Delta \tilde E = {\tilde E_i} - {\tilde E_f} = \frac{{2\beta }}{{1 + \beta }}{{\tilde E}_i}
\end{equation}

The efficiency of a collision $\eta$ can be defined as the fraction of the photon's energy transferred at each reflection. 

\begin{equation}\label{eff1def}
\eta  \equiv \frac{{\Delta \tilde E}}{{{{\tilde E}_i}}} = \frac{{2\beta }}{{1 + \beta }}
\end{equation}

\citet{marx1966interstellar} arrived at the same result in his original paper on interstellar travel. The fact that efficiency approaches unity as $\beta  \to 1$ seems to promise highly efficient, ultra-relativistic travel. However, a detailed analysis of the energy transfer reveals the subtleties involved. 

Energy emitted by the DE system gets transferred into:
\begin{enumerate}
	
	\item forward photon column (FPC) - energy of the photons traveling towards the sail \textit{before} striking the sail
	
	\item the kinetic energy of the spacecraft
	
	\item backward photon column (BPC) - energy of the reflected photons traveling away from the sail \textit{after} striking the sail.
\end{enumerate}

The rate of change of energy in the FPC is: 

\begin{equation}\label{pfpc}
{P_ \rightarrow } = \frac{d}{{dt}}\left( {\frac{\Gamma }{c}x(t){E_0}} \right) = P\beta 
\end{equation}

Using equation \ref{simplediffeqnbeta}, the rate of change of energy of the spacecraft is:

\begin{equation}\label{eff2explicit}
{P_{sc}} = \frac{d}{{dt}}\left( {\gamma m{c^2}} \right) = m{c^2}\dot \beta \beta {\gamma ^3} = 2P\beta \left( {\frac{{1 - \beta }}{{1 + \beta }}} \right)
\end{equation}	 

The rate of change of energy in the BPC can be calculated by viewing the reflected photons as being emitted at a rate $\Gamma\left(1-\beta\right)$ and having a Doppler shifted energy given by equation \ref{doppler}.

\begin{equation}\label{pbpc}
{P_ \leftarrow } = \Gamma \left( {1 - \beta } \right)\left( {\frac{{1 - \beta }}{{1 + \beta }}} \right){E_0} = P\frac{{{{\left( {1 - \beta } \right)}^2}}}{{\left( {1 + \beta } \right)}}
\end{equation}

These results are consistent with energy conservation which demands: ${P_ \leftarrow } + {P_{sc}} + {P_ \rightarrow } = P$. Note that all these powers are calculated in the rest frame of the DE system. We define dimensionless powers $\pi$: 

\begin{equation}
{\pi _i} = \frac{{{P_i}}}{P} \qquad \text{where } i \in \left\{ \rightarrow, sc, \leftarrow \right\}
\end{equation}

\begin{figure}
	\centering
	\includegraphics[width=8cm]{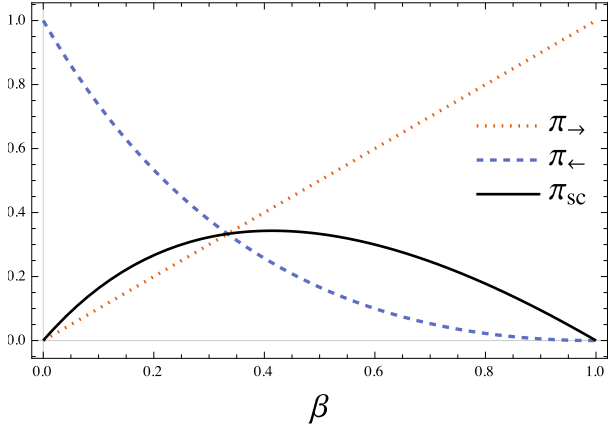}
	\caption{A plot of the various powers in equations \ref{pfpc}, \ref{eff2explicit} and \ref{pbpc}. Initially, the spacecraft velocity is low and Doppler effects are small. Hence, the power in the backward photon column (BPC) dominates. As the spacecraft speeds up, the power on the sail decreases and the reflected photons get increasingly redshifted. This causes the power in the BPC to decrease and the power in the forward photon column (FPC) to increase. At low velocities, spacecraft power increases since the increased efficiency per collision (Eq \ref{eff1def}) outweighs the decreasing flux on the sail. At higher speeds, the decreasing flux dominates and the spacecraft power drops to 0. At $\beta =\sqrt{2}-1$ these competing effects maximize the spacecraft power. At $\beta = 1/3$, the powers in all three components are equal. For large $\beta$, the DE system's power goes into elongating the FPC instead of propelling the spacecraft. Diffraction effects are ignored here but discussed in section \ref{sect:diffraction}}
	\label{powerplots}
\end{figure}

Fig \ref{powerplots} shows how these various powers change with the spacecraft velocity. \citet{marx1966interstellar} and equation \ref{eff1def} correctly conclude that as speed increases, individual photons are more efficient at transferring energy to the spacecraft. However, as speed increases, the photons also take an increasingly long time to catch up to the sail. As a result, at high speeds, energy from the DE system goes into the forward photon column as opposed to accelerating the spacecraft.

\section{Sail of finite size}
\label{sect:diffraction}

\begin{figure}
	\centering
	\includegraphics[scale=0.4]{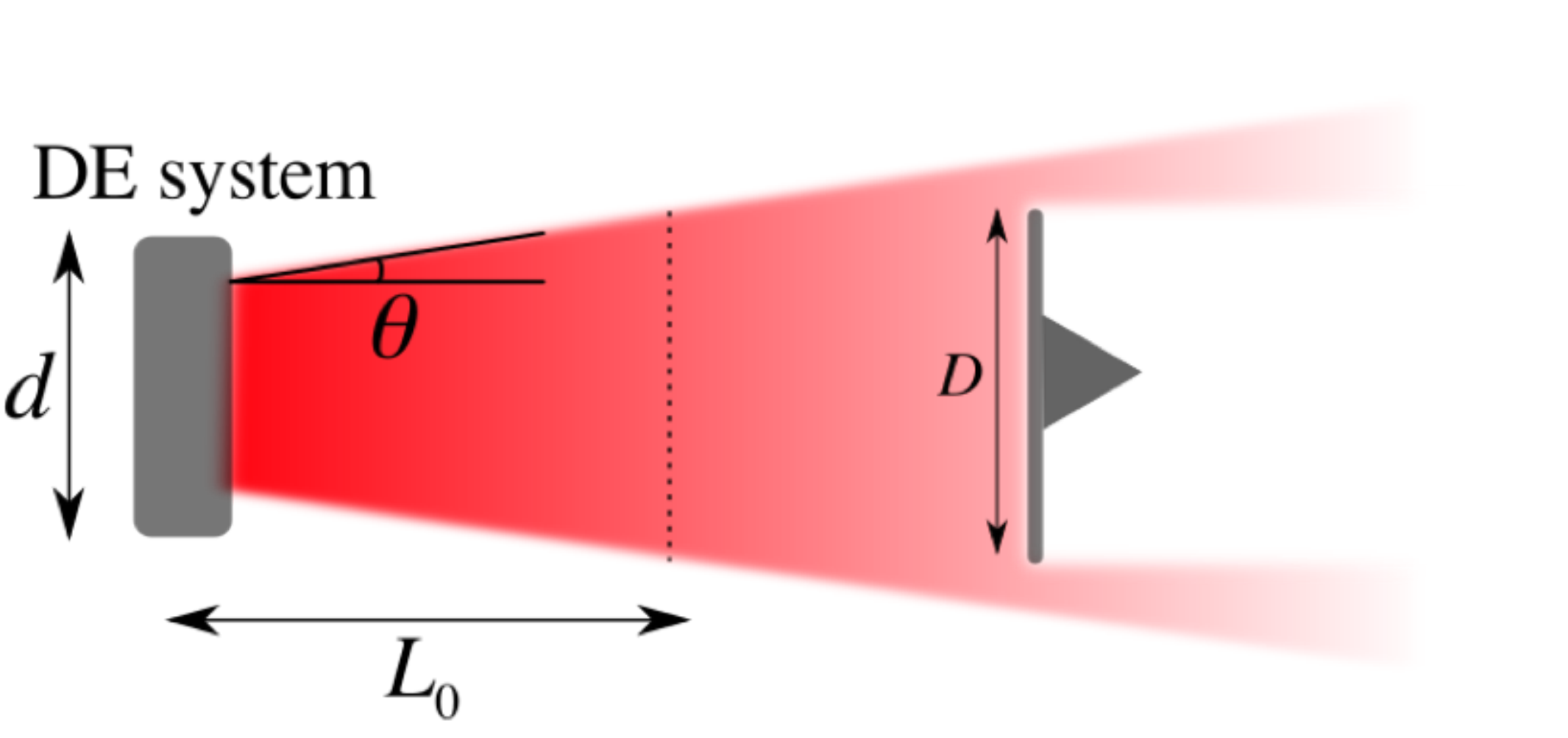}
	\caption{A schematic showing the diffraction of the DE system. When the spacecraft is far from the DE system, the beam spills beyond the sail. $d$ refers to the size of the DE system, $D$ refers to the size of the sail. $\theta$ is the beam divergence and $L_0$ is the critical distance at which the sail is just able to capture the entire beam.}
	\label{diffschematic}
\end{figure}

Until now we have implicitly assumed that all the photons emitted by the DE system eventually intercept the sail. However, the DE system has a finite beam divergence and at large distances, the beam spills far beyond what the area of the sail can capture (figure \ref{diffschematic}). For a physically realistic treatment, diffraction effects must be considered. 

A square DE system of side length $d =$ 10 km emitting at wavelength $\lambda =$ 1064 nm (Nd:YAG laser) has a beam divergence given by:

\begin{equation}\label{sqbeamdiv}
\theta  = \frac{{\lambda }}{d} \approx {10^{ - 10}} \text{rad}
\end{equation}

In general, for an arbitrarily shaped DE system:

\begin{equation}\label{beamdiv}
\theta  = \frac{{{\alpha}\lambda }}{d}
\end{equation} 

The two main cases of interest are $\alpha=1$ for a square DE system and $\alpha=1.22$ (determined by first zero of the Bessel function of the first kind) for a circular DE system.

If these spacecraft are expected to reach distances $\sim 1$ light year, the sail size to catch the entire DE system beam would need to be $\sim 10^6$ m. This is unrealistic and the situation where the DE system's spot size is larger than the sail must be considered. Each spacecraft-system comprises of a sail of mass $m_s$ attached to a bare spacecraft of mass $m_0$. Given the rate at which technology is improving, it is modest to assume a sail with thickness 1 $\mu$m  made out of a material with density 1400 kg m$^{-3}$. Such a sail with a mass of 1 g, would be $0.85$ m on a side. Let $L_0$ be the maximum distance at which the spot size of the DE system fully illuminates a sail of side length $D$. 

\begin{equation} \label{diffractionrelation}
{L_0} = \frac{{dD}}{{2\lambda \alpha }}
\end{equation}

\textit{At any distance $x>L_0$, photons spill beyond the area of the sail according to the inverse square law, reducing incident power on the sail.} This quadratic reduction in incident power leads to a terminal velocity of the spacecraft. Diffraction effects modify equation \ref{simplediffeqnbeta} to equation \ref{diffractiondiffyq}.

\begin{equation}\label{diffractiondiffyq}
\displaystyle
\dot \beta  = \left\{ {\begin{array}{*{20}{c}}
	\displaystyle{\frac{{2P}}{{m{c^2}{\gamma ^3}}}\left( {\frac{{1 - \beta }}{{1 + \beta }}} \right)}&{x < {L_0}}\\
	\displaystyle{\frac{{2P}}{{m{c^2}{\gamma ^3}}}\left( {\frac{{1 - \beta }}{{1 + \beta }}} \right){{\left( {\frac{{{L_0}}}{x}} \right)}^2}}&{x > {L_0}}
	\end{array}} \right.
\end{equation}

When $x<{L_0}$, the acceleration depends only upon the velocity and a closed form solution of $t(\beta)$ can be obtained. When $x>{L_0}$, the acceleration depends on both the velocity and the position. This results in a non-linear, second-order differential equation and finding a closed-form solution for $x(t)$ is difficult. Instead, it is useful to find how the velocity varies with distance by rewriting $\dot v = \left({{dv}}/{{dx}}\right)\left({{dx}}/{{dt}}\right)$ and integrating to find a relationship between $v$ and $x$. 

\subsection{Terminal velocity in non-relativistic limit}

In the non-relativistic limit, equation \ref{diffractiondiffyq} reduces to:
\begin{equation}
m\dot v = \left\{ {\begin{array}{*{20}{c}}
	\displaystyle{\frac{{2P}}{c}}&{x < {L_0}}\\
	\displaystyle{\frac{{2P}}{c}{{\left( {\frac{{{L_0}}}{x}} \right)}^2}}&{x > {L_0}}
	\end{array}} \right.
\end{equation}

Integrating to find $v(x)$ in the two different regimes yields:

\begin{equation}
v\left( x \right) = \left\{ {\begin{array}{*{20}{c}}
	\displaystyle{\sqrt {\frac{{4P}}{{mc}}x} }&{x < {L_0}}\\
	\displaystyle{\sqrt {\frac{{4PL_0^2}}{{mc}}\left( {\frac{2}{{{L_0}}} - \frac{1}{x}} \right)} }&{x > {L_0}}
	\end{array}} \right.
\end{equation}

At infinity, the maximum velocity that can be reached is given by:

\begin{equation}\label{nrterminal}
v\left( \infty  \right) = {v_{\max }} = \sqrt {\frac{{8P{L_0}}}{{mc}}}  = \sqrt 2 {v\left(L_0\right)}
\end{equation}

For massive spacecrafts that don't reach relativistic speeds, equation \ref{nrterminal} is a good approximation. By the time the spacecraft reaches $x=L_0$, it's already traveling at over $70\%$ of it's terminal velocity. Optimizing spacecraft speed is thus a matter of maximizing the speed attained before diffraction effects start becoming relevant. 

\subsection{Optimizing spacecraft design}

We're interested in finding the optimum sail size to maximize the speed of the spacecraft at $L_0$. On one hand, the bigger the sail, the more time the spacecraft spends in the $x<L_0$ regime. However, a large sail is also heavy so the spacecraft's acceleration reduced. 

Using the same chain rule as before, we integrate equation \ref{diffractiondiffyq} to find a relationship between $v$ and $x$ in the $x<L_0$ regime. In particular, at $x=L_0$, we have:

\begin{equation}\label{maxbeta}
\frac{{2{\beta _0} - 1}}{{3{\gamma _0}{{\left( {1 - {\beta _0}} \right)}^2}}} = \frac{{2P}}{{m{c^3}}}{L_0} - \frac{1}{3}
\end{equation}

The left hand side of equation \ref{maxbeta} monotonically increases with $\beta_0$. Thus, maximizing $\beta_0$ is equivalent to maximizing the right hand side of equation \ref{maxbeta}. The total mass of a spacecraft-system is $m = m_0 + m_s$. For a sail of size $D$ with a fixed thickness $h$ and density $\rho$, ${m_s} = k\rho {D^2}h$. $k$ is a dimensionless constant depending on the geometry of the sail ($k=1$ for a square sail,  $k=\pi/4$ for a circular sail).  Maximizing the right hand side of equation \ref{maxbeta} with respect to $D$ yields:  $k\rho {D^2}h = {m_o} = {m_s}$. Thus, the spacecraft reaches it's maximum speed when the \textit{mass of the sail equals the mass of the spacecraft.} 

\subsection{Possible mission scenarios}

Under the optimum condition of $m_s = m_0$ several possible mission scenarios for propelling various spacecrafts can be considered. Table \ref{missionscenarios} lists a variety of missions that span 8 orders of magnitude in spacecraft mass. Although DE propulsion allows low mass spacecraft to achieve relativistic speeds, it is also useful for interplanetary missions in larger spacecraft. For instance, a $10^5$ kg spacecraft capable of containing a manned crew attains a terminal velocity of $0.002c$. The versatility and scalability of these missions make DE propulsion an attractive prospect for space exploration. With advances in materials science and nanotechnology, it's possible to envision sails $\lesssim 10^{-2}$ $\mu$m thick. For a given mass, the thinner the sail, the larger it can be.  Consequently, the spacecraft spends more time in the $x<L_0$ regime and achieves a higher speed (table \ref{sail thickness}).

\begin{table*}
\centering		
		\begin{tabular}{| c | c | c | c | c | c | c |} 
			\hline
			Sail Thickness ($\mu$m)  & Spacecraft mass (kg) & Sail size (m) & $t_0$ (s) & $L_0$ (m) & $\beta_0$ & $\beta_{max}$ \\  [1ex] 
			\hline
			\hline
			1.0 & $10^{-3}$ & 0.85 & 164 & $4.0 \times {10^9}$ & 0.15 & 0.21 \\  [1ex]  				
			\hline
			1.0 & $10^{-2}$ & 2.7 & 901 & $1.3 \times {10^{10}}$ & 0.091 & 0.13 \\  [1ex]
			\hline
			1.0 & $10^{-1}$ & 8.5 & $4.99 \times 10^3$ & $4.0 \times {10^{10}}$ & 0.053 & 0.073 \\  [1ex]  
			\hline
			1.0 & $10^{0}$ & 27 & $2.79 \times 10^4$ & $1.3 \times {10^{11}}$ & 0.030 & 0.042 \\  [1ex]  
			\hline
			1.0 & $10^{1}$ & 85 & $1.56 \times 10^5$ & $4.0 \times {10^{11}}$ & 0.017 & 0.024 \\  [1ex]  
			\hline
			1.0 & $10^{2}$ & 270 & $8.75 \times 10^5$ & $1.3 \times {10^{12}}$ & 0.0096 & 0.014 \\  [1ex]  
			\hline
			1.0 & $10^{3}$ & 850 & $4.91 \times 10^6$ & $4.0 \times {10^{12}}$ & 0.0054 & 0.0077 \\  [1ex]  
			\hline
			1.0 & $10^{4}$ & 2700 & $2.76 \times 10^7$ & $1.3 \times {10^{13}}$ & 0.0031 & 0.0043 \\  [1ex]  
			\hline
			1.0 & $10^{5}$ & 8500 & $1.55 \times 10^8$ &$4.0 \times {10^{13}}$ & 0.0017  & 0.0024 \\  [1ex]  
			\hline
			
		\end{tabular}

	\caption{
		 Mission scenarios for a 100 GW (10 km) DE system propelling various bare spacecraft masses. We assume the optimal case where $m_s = m_0$ and a square sail with thickness 1 $\mu$m  made out of a material with a density of 1400 kg m$^{-3}$. $L_0$ is the distance at which the sail size equals the DE system's diffraction spot size. $t_0$ is the time taken to accelerate from rest to a distance $L_0$. $\beta_0$ is the speed of the spacecraft at $L_0$ and $\beta_{max}$ is the spacecraft's terminal speed.}
	\label{missionscenarios}

\end{table*}

\begin{table*}
	
	\begin{center}
		\begin{tabular}{| c | c | c | c | c | c | c |} 
			\hline
			Sail thickness ($\mu$m) & Spacecraft mass (kg) & Sail size (m) & $t_0$ (s) & $L_0$ (m) & $\beta_0$ & $\beta_{max}$ \\  [1ex] 
			\hline
			\hline
			$0.01$ & $10^{-3}$ & 8.5 & 587 & $4.0 \times {10^{10}}$ & 0.39 & 0.48 \\  [1ex]  				
			\hline
			$0.05$ & $10^{-3}$ & 3.8 & 370 & $1.8 \times {10^{10}}$ & 0.29 & 0.37 \\  [1ex]
			\hline
			$0.1$ & $10^{-3}$ & 2.7 & 306 & $1.3 \times {10^{10}}$ & 0.25 & 0.33 \\  [1ex]  
			\hline
			$0.5$ & $10^{-3}$ & 1.2 & 198 & $5.7 \times {10^{9}}$ & 0.18 & 0.24 \\  [1ex]  
			\hline

		\end{tabular}
	\end{center}
	\caption{ Mission scenarios for a 1 g bare spacecraft under constant illumination from a 100 GW (10 km) DE system. We assume the optimal case where $m_s = m_0$ and a square sail with thickness 1 $\mu$m made out of a material with a density of 1400 kg m$^{-3}$. We explore the effect of reducing sail thickness on various mission parameters.}
    
	\label{sail thickness}
\end{table*}

\begin{figure}
	\centering
	\includegraphics[width=8cm]{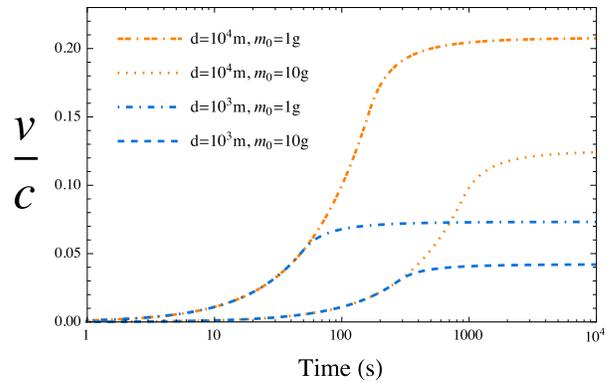}
	\caption{Velocity curves for different bare spacecraft masses ($m_0$ = 1 g, 10 g) and DE system sizes ($d$ = 1 km, 10 km). We assume a 100 GW DE system and the optimal case where $m_s =m_0$. The sail is assumed to have thickness 1 $\mu$m  and density 1400 kg m$^{-3}$. The 1 km DE system has a larger beam divergence than the 10 km system. As a result, spacecrafts propelled by the 1 km DE system spend less time in the $x<L_0$ regime and typically have lower terminal velocities. Gram-scale spacecrafts accelerate extremely quickly reaching their relativistic terminal velocity in a matter of minutes.}
	\label{starshotplots}
	
\end{figure}

\begin{figure}
	\centering
	\includegraphics[width=8cm]{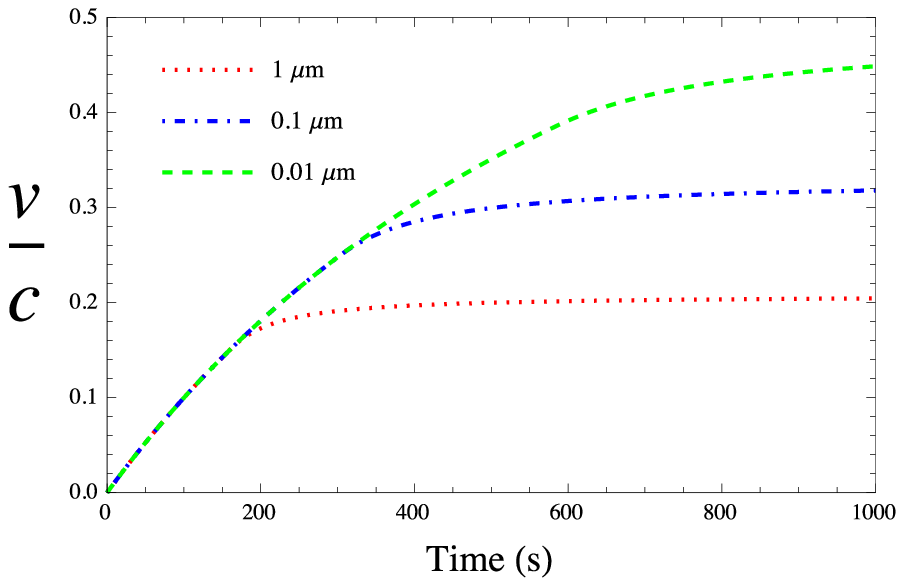}
	\caption{Velocity curves for a 1 g bare spacecraft with varying sail thicknesses. We assume the optimal case where $m_s=m_0$ and a 10 km, 100 GW DE system. The three trajectories are identical early on when they are all in the $x<L_0$ regime and are completely determined by equation \ref{vclosedformt}. As each sail subsequently moves into the $x>L_0$ regime, diffraction effects become important and the spacecraft quickly approaches a terminal speed. Thin sails $\sim 0.01 \mu$m allow for highly relativistic travel at speeds $\sim 0.5 c$. }
	\label{sailthickness}
\end{figure}

Solving equation \ref{diffractiondiffyq} numerically gives $v(t)$ for a set of parameters. The results are shown in figures  \ref{starshotplots} and \ref{sailthickness}.

Figure \ref{starshotplots} shows that gram-scaled spacecrafts can be rapidly accelerated to relativistic speeds. By further optimizing parameters such as using a larger, more powerful DE system with lighter spacecraft and thinner sails, it is possible to reach highly relativistic speeds. In figure \ref{sailthickness}, sails with thicknesses of 1 $\mu$m , $0.1\mu$m and $0.01\mu$m can reach Alpha Centauri in 25 yrs, 15 yrs and 10 yrs respectively (after which it would take 4 years for Earth to receive any transmitted information). Given these short time scales, one can imagine interstellar exploration using millions of tiny spacecraft propelled one after another by DE systems. 

\section{Photon Recycling}
\label{sect:pr}

\textit{Photon recycling} involves letting the photons repeatedly reflect between the DE system and the sail in order to maximize the energy transfer to the spacecraft. To analyze an ideal case, we assume the DE system and sail act as retro-reflectors so that any incident light is reflected back to the source. Even so, diffraction effects still need to be considered. This is a complex problem since at any given time, there are a range of photons with different frequencies striking the sail. As a first approximation, we perform a non-relativistic analysis while still retaining the key physics.

\subsection{Non-relativistic approximation}

First, consider the regime where $x<L_0$. Let ${D_S}\left( N \right)$ be the spot diameter on the sail at the $N$th reflection on the sail. Upon emission from the DE system, light diffracts and the spot size at a distance $x$ is given by:

\begin{equation}
D_S\left( 1 \right) = \frac{2\lambda}{d}x
\end{equation}

Light now diffracts out of the sail and back towards the DE system. Let ${d_A}\left( N \right)$ be the spot size on the DE system at the $N$th reflection on the DE system:

	\begin{equation}
	{d_A}\left( 1 \right) = \frac{{2\lambda }}{{{D_S}\left( 1 \right)}}{x} = d
	\end{equation} 
    
    Subsequent reflections keep propagating between the DE system and the sail such that:
	
	\begin{equation}
	{D_S}\left( N \right) = \frac{{2\lambda }}{{{d_A}\left( {N - 1} \right)}}{x} = \frac{{2\lambda {x}}}{d}
	\end{equation}
	
	\begin{equation}
	{d_A}\left( N \right) = \frac{{2\lambda }}{{{D_S}\left( {N - 1} \right)}}{x} = d
	\end{equation}

In the $x<L_0$ regime, spot sizes on the sail and DE system will remain within their respective sizes i.e. there is no leakage due to overflowing photons. However, compounded over multiple reflections, the loss due to the reflectivity's of the DE system and sail become significant. Let the reflectivity of the sail and the DE system be $\alpha_1$ and $\alpha_2$ respectively. Between each reflection at the sail, the power reduces by a factor of $\alpha_1\alpha_2$. This system can be treated as a new DE system with power $P_r$ where $P_r$ arises from a sum over all reflections.

\begin{equation}\label{recyclingclose}
	{P_r} = P\left[ {1 + \left( {{\alpha _1}{\alpha _2}} \right) + {{\left( {{\alpha _1}{\alpha _2}} \right)}^2} + ...} \right] = \frac{P}{{1 - {\alpha _1}{\alpha _2}}}
	\end{equation}

In the $x>L_0$ regime, there will be additional power loss since photons diffract past the DE system and sail and 'leak' off into space. For distances larger than $L_0$, at each bounce there will be another loss factor depending on the fractions of the DE system and sail that are able to reflect the incoming beam. The fraction of the beam that reaches the sail $f_S$ equals:
	
	\begin{equation}
    {f_S} = {\left( {\frac{D}{{{D_{beam}}}}} \right)^2} = {\left( {\frac{{Dd}}{{2\lambda x}}} \right)^2} = {\left( {\frac{{{L_0}}}{x}} \right)^2}
	\end{equation}
	
    Similarly, the fraction of the beam that reaches back to the DE system $f_A$ equals:
    
    \begin{equation}
    {f_A} = {\left( {\frac{d}{{{D_{beam}}}}} \right)^2} = {\left( {\frac{{Dd}}{{2\lambda x}}} \right)^2} = {\left( {\frac{{{L_0}}}{x}} \right)^2}
    \end{equation}
 
Combining the two results, a factor of ${f_A}{f_R}{\alpha_1}{\alpha_2}$ is lost between consecutive reflections on the sail. Thus when $x>L_0$, equation \ref{recyclingclose} must be modified to :	
  
  \begin{equation}\label{recyclingfar}
  \begin{split}
  {P_r} &= P\left[ {1 + \left( {{f_a}{f_s}{\alpha _1}{\alpha _2}} \right) + {{\left( {{f_a}{f_s}{\alpha _1}{\alpha _2}} \right)}^2} + ...} \right]\\ &= \frac{P}{{1 - {\alpha _1}{\alpha _2}{{\left( {\frac{{{L_0}}}{x}} \right)}^4}}}	
  \end{split}
  \end{equation}
In general:

\begin{equation}\label{prpower}
\displaystyle
{P_r} = \left\{ {\begin{array}{*{20}{c}}
	\displaystyle{\frac{P}{{1 - {\alpha _1}{\alpha _2}}}}&{x < {L_0}}\\
	\displaystyle{\frac{P}{{1 - {\alpha _1}{\alpha _2}{\left( {\frac{{{L_0}}}{x}} \right)^4}}}}&{x > {L_0}}
	\end{array}} \right.
\end{equation}

\begin{figure}
	
	\centering
	\includegraphics[width=8cm]{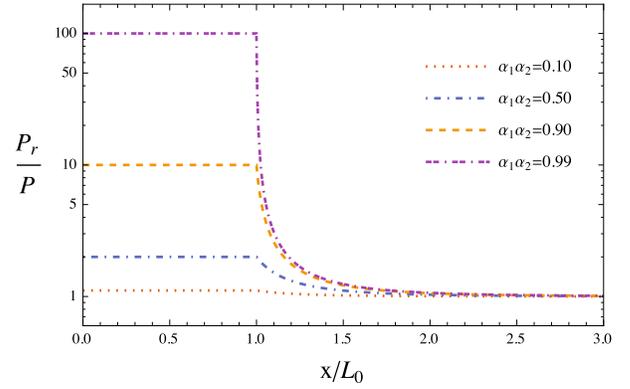}
	\caption{Additional power gained by photon recycling as a function of the distance between the DE system and the sail. Plotted curves correspond to different DE system-sail reflectivity's. Photon recycling is beneficial when $x<L_0$. For $x>L_0$ diffraction losses quickly dominate and very little additional power is gained.}

	\label{prpowerfig}
\end{figure}

Figure \ref{prpowerfig} shows the power gained by photon recycling at various distances and reflectivity's.

We note that equation \ref{prpower} only provides an approximation i.e. by the time the spacecraft travels a distance $\sim L_0$, our non-relativistic approximations may no longer be valid. 

As seen in figure \ref{prpowerfig}, photon recycling is only beneficial when $x<L_0$. Our non-relativistic  calculation gives a rough approximation at early times when the reflection rate is high, $\beta \ll 1 $ and $x\ll L_0$. Incidentally, this is also the regime in which photon recycling is the most effective (Fig \ref{prpowerfig}). 

\subsection{Photon recycling in the relativistic limit}

Our simplified treatment of photon recycling is applicable and beneficial for low speed spacecraft. At higher speeds our analysis breaks down because:

\begin{itemize}
\item the flight time of the photons has been ignored. This is unphysical and a causal, relativistic theory of photon recycling needs to take this into account.
\item energy losses (from Doppler shifts) due to the relative motion between the spacecraft and the DE system have been ignored. Doppler shift is the mechanism by which energy is transferred from the photons to the spacecraft and needs to be included in a rigorous analysis. 
\end{itemize}

Taking both these factors into account significantly complicates the problem. In a given time interval, there will be photons with a range of frequencies striking the sail. The frequency of each photon will depend on the number of reflections it has encountered in the past. In a future work, we will take these factors into account to discuss a relativistic model of photon recycling.

\section{Conclusion}
\label{sect:conclusion}

A relativistic solution of a spacecraft driven by a directed energy system has been derived and applied to the NASA Starlight program.
We address claims made in \citep{kipping} asserting that our previous solution (which we derive in detail here) is incorrect. We resolve this discrepancy by showing that \citep{kipping} fails to include the travel time of the photons and hence violates causality. Diffraction effects of the directed energy system reduce the incident power on the sail at large distances, thus limiting the spacecraft speed. We show that the spacecraft speed is maximized by using a sail mass equal to the spacecraft mass, consistent with the non-relativistic solution in \citep{lubin2016roadmap}. As an example, when propelled by a 100 GW directed energy system, gram-scale spacecrafts can accelerate up to $0.2c$ within a few minutes of illumination (with higher or lower speeds dependent on the system parameters chosen). This opens up the possibility of relativistic flight allowing humanity's first interstellar missions. We discuss the effects of photon recycling and show it is most effective for low speed missions and of limited utility for relativistic missions. We emphasize that this technology is not limited to gram-scale spacecraft and show the scaling of speed with a variety of system parameters. 

\acknowledgments

We gratefully acknowledge funding from NASA NIAC NNX15AL91G and NASA NIAC NNX16AL32G and the NASA California Space Grant NASA NNX10AT93H as well as a generous gift from the Emmett and Gladys W. technology fund in support of this research. PML acknowledges support from the Breakthrough Starshot program.\\

\textbf{Web based calculator} – A web based calculator based on the work in this paper and more details on our NASA Starlight program can be found at
\href{http://www.deepspace.ucsb.edu/projects/starlight}{http://www.deepspace.ucsb.edu/projects/starlight}.

\bibliography{relativisticrefs}
\end{document}